\documentclass[twocolumn]{aastex631}
\usepackage{graphicx}
\usepackage{natbib}
\usepackage{longtable}
\usepackage{appendix}
\newcommand\aastex{AAS\TeX}

\received{2023 February 27}
\revised{2023 June 19}
\accepted{2021 June 24}
\submitjournal{ApJ}

\shorttitle{\aastex\ SFE}
\shortauthors{Lee et al. 2023}

\begin{document}

\title{The Origin of Star Formation in Early-type Galaxies Inferred from Spatially Resolved Spectroscopy}

\correspondingauthor{Ho Seong Hwang}
\email{galaxy79@snu.ac.kr}

\author[0000-0003-2779-6793]{Yun Hee Lee}
\affil{Department of Astronomy and Atmospheric sciences, Kyungpook National University, Daegu 41566, Korea}
\affil{Korea Astronomy and Space Science Institute (KASI), 776 Daedeokdae-ro, Yuseong-gu, Daejeon 34055, Korea}

\author[0000-0003-3428-7612]{Ho Seong Hwang}
\affil{Astronomy Program, Department of Physics and Astronomy, Seoul National University, 1 Gwanak-ro, Gwanak-gu, Seoul 08826, Korea}
\affil{SNU Astronomy Research Center, Seoul National University, 1 Gwanak-ro, Gwanak-gu, Seoul 08826, Korea}
\affil{Korea Astronomy and Space Science Institute (KASI), 776 Daedeokdae-ro, Yuseong-gu, Daejeon 34055, Korea}

\author[0000-0002-2013-1273]{Narae Hwang}
\affil{Korea Astronomy and Space Science Institute (KASI), 776 Daedeokdae-ro, Yuseong-gu, Daejeon 34055, Korea}

\author[0000-0003-0283-8352]{Jong Chul Lee}
\affil{Korea Astronomy and Space Science Institute (KASI), 776 Daedeokdae-ro, Yuseong-gu, Daejeon 34055, Korea}

\author[0000-0001-8638-4460]{Ki-Beom Kim}
\affil{Department of Biology, Teachers College and Institute for Phylogenomics and Evolution, Kyungpook National University, Daegu 41566, Korea}

\begin{abstract}
We investigate the origin of star formation activity in early-type galaxies with current star formation using spatially resolved spectroscopic data from the Mapping Nearby Galaxies at APO (MaNGA) in the Sloan Digital Sky Survey (SDSS). We first identify star-forming early-type galaxies from the SDSS sample, which are morphologically early-type but show current star formation activity in their optical spectra. We then construct comparison samples with different combinations of star formation activity and morphology, which include star-forming late-type galaxies, quiescent early-type galaxies and quiescent late-type galaxies. Our analysis of the optical spectra reveals that the star-forming early-type galaxies have two distinctive episodes of star formation, which is similar to late-type galaxies but different from quiescent early-type galaxies with a single star formation episode. Star-forming early-type galaxies have properties in common with star-forming late-type galaxies, which include stellar population, gas and dust content, mass and environment. However, the physical properties of star-forming early-type galaxies derived from spatially resolved spectroscopy differ from those of star-forming late-type galaxies in the sense that the gas in star-forming early-type galaxies is more concentrated than their stars, and is often kinematically misaligned with stars. The age gradient of star-forming early-type galaxies also differs from those of star-forming late-type galaxies. Our findings suggest that the current star formation in star-forming early-type galaxies has an external origin including galaxy mergers or accretion gas from the cosmic web.

\end{abstract}
\keywords{galaxies: elliptical and lenticular, cD -- galaxies: star formation -- galaxies: evolution -- galaxies: structure}

\section{Introduction}\label{chap1}
Understanding the physical mechanism behind the bimodal distribution of galaxies is one of key issues in the study of galaxy evolution: red elliptical galaxies and blue spiral galaxies, which are called red sequence and blue cloud, respectively \citep[e.g.][]{2007Faber}. They are distributed differently in various parameter spaces including the color - magnitude diagram and the star formation rate (SFR) - mass diagram that can be derived from photometric and spectroscopic surveys, respectively \citep{2001Strateva, 2004Baldry, 2004Bell, 2007Faber, 2012Wetzel, 2019Liu}. The red sequence galaxies are generally more massive, luminous, bulge-dominated, and composed of older stellar populations. On the other hand, the blue cloud galaxies are less massive, less luminous, disk-dominated, and composed of younger stellar populations. They are mainly understood as the evolutionary sequence from star-forming spiral galaxies to quiescent elliptical galaxies through the morphological transformation and star formation quenching \citep{2008Park, 2009Hwang, 2021Liu}. These are thought to be driven by the processes such as galaxy mergers and AGN feedback \citep{2006Croton, 2007Faber, 2010Hopkins}.

On the other hand, the discovery of quiescent/ red late-type galaxies \citep[q-LTGs,][]{2010Bundy} and star-forming/ blue early-type galaxies \citep[SF-ETGs,][]{2004Fukugita} could suggest more diverse pathways of galaxy evolution \citep{2022Park} or provide hints for the processes of morphological transformation and of star formation quenching \citep{2021Liu}. SF-ETGs show strong H$\alpha$ emission similar to late-type galaxies, indicating active star formation on the timescale of $10^6-10^7$yr for OB stars \citep{2004Fukugita, 2012Kennicutt}. They are rare populations, accounting for 4\% out of early-type galaxies in the local universe at z $<$ 0.12 \citep{2019Liu}. They are less massive and luminous, which include both new and old stellar populations unlike quiescent early-type galaxies \citep[q-ETGs,][]{2009Huang, 2021Liu}. They were suggested to be a progenitor of q-ETGs \citep{2004Fukugita} or the outcome of rejuvenation of q-ETGs \citep{2009Huang}. They could be also in the middle stage of the morphological transformation by merger and the star formation quenching by AGN \citep{2022Park}. In this paper, we aim to figure out the origin of SFEs, in particular, by exploring their spatially resolved properties with the Integral Field Spectroscopy (IFS) data from the Mapping Nearby Galaxies at APO \citep[MaNGA;][]{2015Bundy} in the Sloan Digital Sky Survey \citep[SDSS;][]{2000York}.


Interestingly, there are blue early-type galaxies, which are similar to SF-ETGs in terms of their physical properties, but are chosen based on their colors rather than their star formation rates \citep{2006Lee, 2009Schawinski, 2010Huertas}. These galaxies tend to be less luminous and less massive than normal early-type galaxies \citep{2006Lee, 2010aLee}. However, it is important to note that they may not always be consistent with SF-ETGs because optical colors are sensitive to a longer timescale of $10^9$yr \citep{2014Schawinski}. Indeed, \citet{2009Schawinski} used the sample of Galaxy Zoo and found that only 50\% out of blue early-type galaxies show active star formation.

In this work, we use a sample of galaxies at $z<0.15$ from SDSS/DR17 \citep{2022Abdurrouf} and classify them into four types based on their morphology and the positions in the SFR - mass diagram: star-forming early- (SF-ETG) and late-type (SF-LTG) galaxies, quiescent early- (q-ETG) and late-type (q-LTG) galaxies \citep{2021Liu}. We compare the stellar population, gas properties, and environments of the four types of galaxies. In particular, we examine the spatially resolved properties of the stellar age, metallicity, velocity, and H$\alpha$ emission for the galaxies in the MaNGA survey. This paper is organized as follows. Section \ref{chap2} describes our sample and the classification. We analyze the stellar population, gas properties, and environment in Sections \ref{chap3}, \ref{chap4}, and \ref{chap5}, in turn. Sections \ref{chap6} and \ref{chap7} are assigned to discussion and summary, respectively.


\begin{figure*}[htb]
\centering
\includegraphics[bb = 30 640 320 780, width = 0.77 \linewidth, clip=]{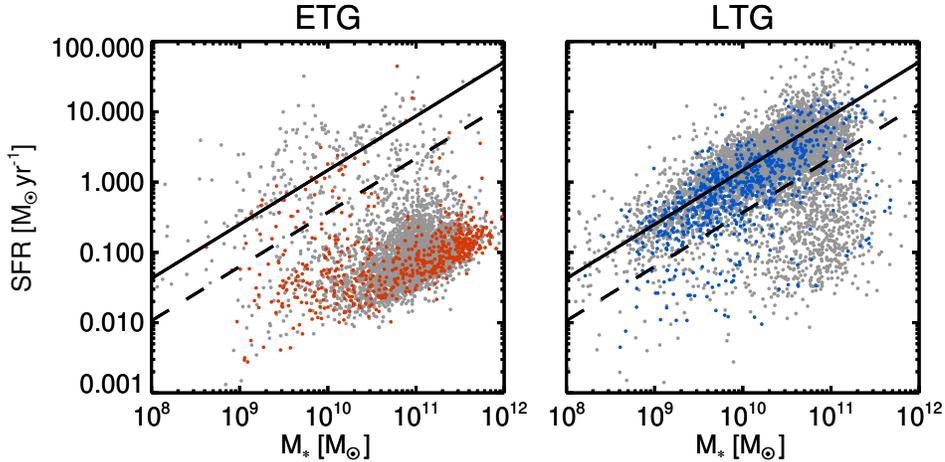}
\caption{Distributions of SDSS galaxies on the SFR - mass diagram. The left and right panels show the distribution of early- and late-type galaxies, respectively. Only 2\% of the data is shown in the diagram for clarity. The red and blue circles represent 20\% of galaxies with IFS data from the MaNGA survey. The thick solid line on the diagram represents the star-forming main-sequence \citep{2007Elbaz} and the dashed line represents a factor of four below the star-forming main-sequence. We classify galaxies into star-forming and quiescent galaxies by the criterion of the dashed line.}\label{Fig1}
\end{figure*}

\section{Samples and Classification} \label{chap2}

To study the properties of SF-ETGs in the local Universe, we start from a sample of $\sim$ 580,000 including early types (i.e. ellipticals and lenticulars) and late types  at $z < 0.15$ from SDSS/DR17. We utilize the galaxy morphological information from \citet{2018Lee}, which combines data of the Korea Institute for Advanced Study (KIAS) Data Release 7 (DR7) Value-Added Galaxy Catalog \citep[VAGC;][]{2010Choi}, the Galaxy Zoo Catalogs \citep{2011Lintott, 2013Willett}, and additional visual inspections \citep{2018Lee}. The KIAS DR7 VAGC provides morphological classification by primarily considering the position in the $(u-r)$ color versus $(g-i)$ color gradient space, while also taking into account the $i$-band concentration index \citep{2005Park}. Subsequently, multiple experienced astronomers conducted an additional visual check of SDSS color images for galaxies with ambiguous automated classifications to ensure final confirmation. The classification process demonstrates a completeness and reliability rate of over 88\% for SDSS galaxies with $m_r < 17.5$. The Galaxy Zoo Catalogs involve the morphological classification of SDSS galaxies based on visual inspections by numerous anonymous citizens \citep{2011Lintott, 2013Willett}. Comparing the KIAS DR7 VAGC and the Galaxy Zoo Catalogs, there is an agreement about 81 per cent for galaxies with $m_r < 17.77$ mag \citep{2018Lee}. When the two catalogs provide different classifications, we prioritize the morphological information from the KIAS DR7 VAGC. A detailed understanding of the difference between the two catalogs is beyond the scope of the paper. Additionally, for 2367 galaxies with $m_r < 17.77$ mag that are not included in either catalog, the authors of \citet{2018Lee} conducted visual inspections.

We also classify the sample galaxies into two groups (i.e. star-forming and quiescent galaxies) according to their positions relative to the star-forming main sequence in the plane of SFR and stellar mass. We adopt the equation for the star-forming main sequence from \citet{2007Elbaz}, which is followed as
\begin{eqnarray}\label{eq1}
SFR^{z\sim0}_{SDSS}[M_{\odot}\rm yr^{-1}]=8.7\times[M_{\ast}/10^{11}M_\odot]^{0.77}. 
\end{eqnarray}
Figure \ref{Fig1} displays the SFR - mass diagram for 173,349 early- and 260,788 late-type galaxies after excluding AGN-host galaxies (i.e. spectral types of composite, AGN, and low S/N LINER from the MPA/JHU VAGC) whose emission line fluxes may not be directly related to star formation \citep{2004Brinchmann}. The SFRs and stellar mass estimates are adopted from the MPA/JHU VAGCs \citep{2004Brinchmann}. For fair comparison with \citet{2007Elbaz}, we convert SFR and stellar mass calculated for Kroupa initial mass function \citep[IMF;][]{2001Kroupa} to Salpeter IMF \citep{1955Salpeter} by dividing them by a factor of 0.7 following \citet{2007Elbaz}.  We define star-forming and quiescent galaxies as those above and below the dashed line, respectively; the dashed line is the one with SFRs four times lower than the solid line for the star-forming main sequence \citep{2017Kim}. The SFR is derived from the SDSS fiber spectrum with the extinction and aperture correction. The stellar mass is calibrated by fitting the spectral energy distribution (SED) to the SDSS $ugriz-$band photometry with the stellar population model \citep{2003Bruzual}. 

\begin{figure*}[htb]
\centering
\includegraphics[bb = 50 12 370 480, width = 0.80 \linewidth, clip=]{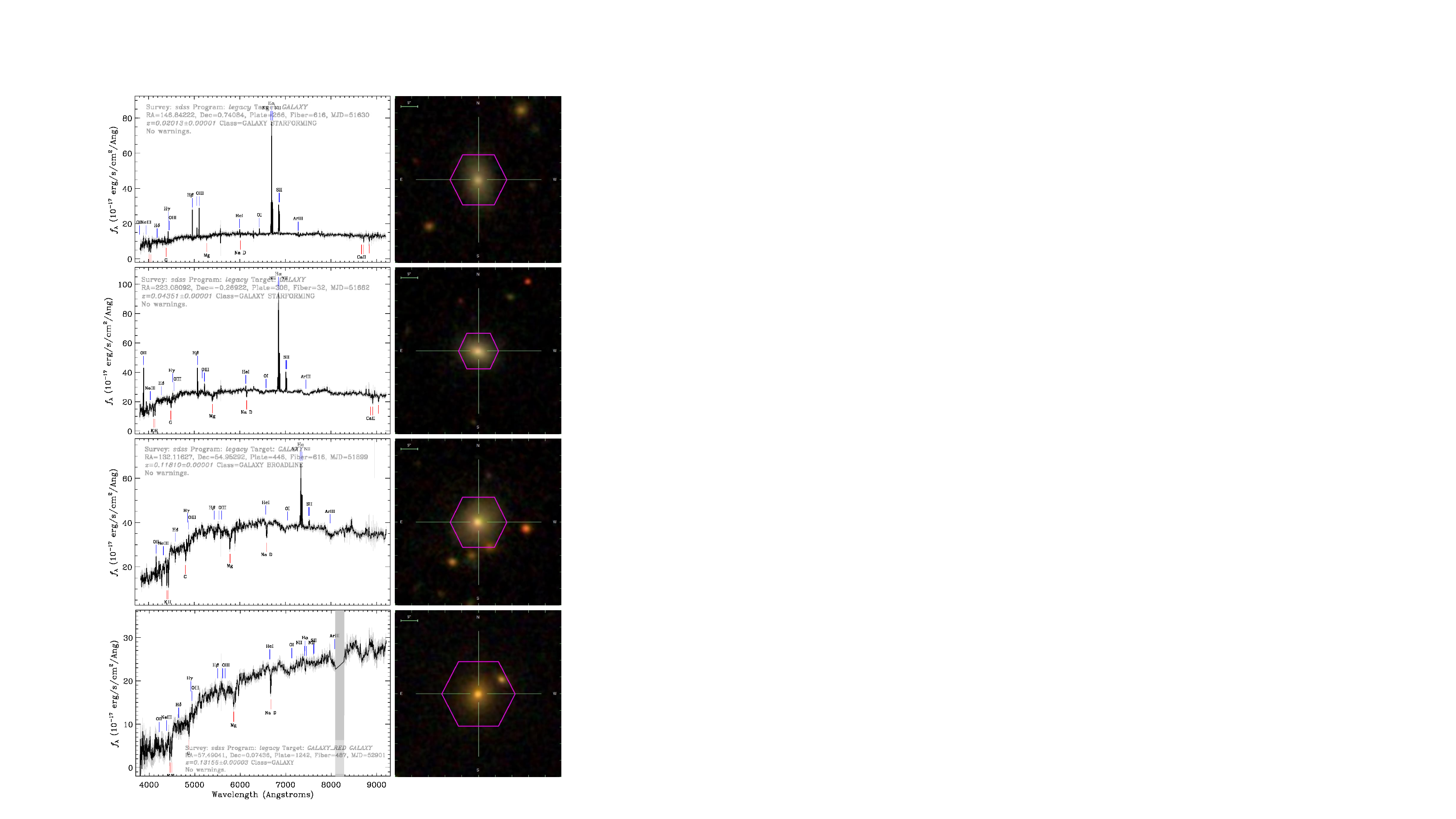}
\caption{Example spectra (left) and images (right) of SF-ETGs (top three rows) and q-ETGs (bottom row). The spectra and images are sourced from SDSS/DR17. The three SF-ETGs are arranged in ascending order of stellar mass: above $10^9\rm M_{\odot}$, $10^{10}\rm M_{\odot}$, and $10^{11}\rm M_{\odot}$, from top to bottom. The example galaxy of q-ETG is selected with a stellar mass exceeding $10^{11}\rm M_{\odot}$, which falls within the common mass range of q-ETGs.}\label{Fig2}
\end{figure*}

We find that 7\% (11,365) of early-type galaxies are classified as star-forming galaxies (left panel in Figure \ref{Fig1}), while 78\% (202,361) of late-type galaxies correspond to star-forming galaxies (right panel in Figure \ref{Fig1}). We overlay the galaxies that have IFS data from MaNGA/DR17 by red and blue circles. These are 2732 early-type galaxies with 191 (7\%) SF-ETGs and 2541 (93\%) q-ETGs (left panel), and 3692 late-type galaxies with 2924 (79\%) SF-LTGs and 768 (21\%) q-LTGs (right panel). Figure \ref{Fig2} shows example spectra (left panel) and images (right panel) of galaxies classified as SF-ETG (top three rows) and q-ETG (bottom row). The SF-ETGs show distinct emission lines including $H\alpha$ in their spectra, unlike q-ETGs.

\begin{figure*}[htb]
\centering
\includegraphics[bb = 10 640 320 780, width = 0.77 \linewidth, clip=]{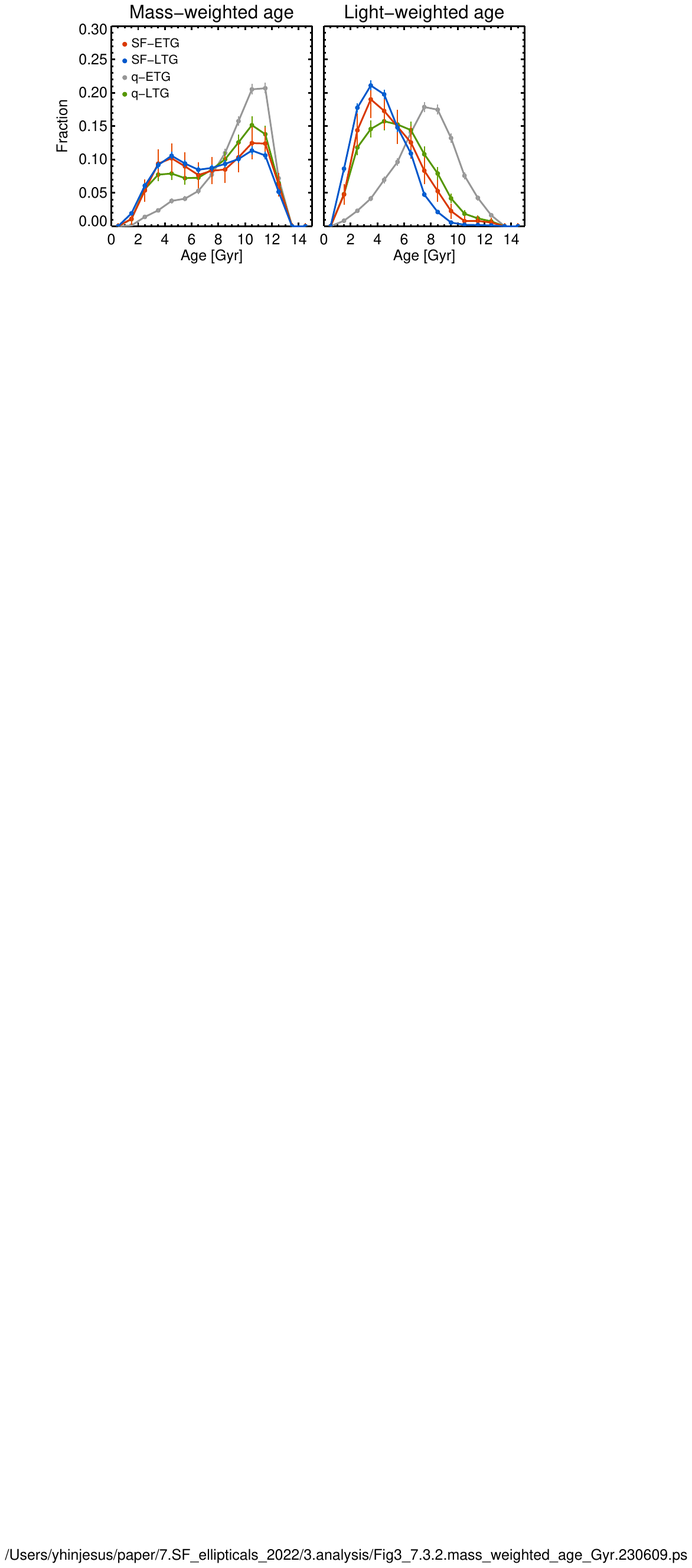}
\caption{Fractions of different stellar populations in the total stellar mass. The four different types of galaxies, divided by their morphology and star-forming status, are represented by different colored lines: star-forming early- (SF-ETG) and late-type (SF-LTG) galaxies, and quiescent early- (q-ETG) and late-type (q-LTG) galaxies are shown in red, blue, gray, and green, respectively. The ages are estimated using two different methods: mass-weighted (left) and light-weighted (right).  The errors are estimated from the confidence interval of the fraction for a beta distribution.} \label{Fig3}
\end{figure*}

\section{Stellar Population} \label{chap3}

We use the MaNGA FIREFLY Value-Added-Catalog \citep{2022Neumann} to investigate the stellar populations. The catalog provides spatially resolved information on the stellar age, metallicity, mass, star formation history, and dust attenuation for the final sample of 10,010 galaxies in the MaNGA survey. The catalog was generated using the full spectral fitting code FIREFLY \citep[Fitting IteRativEly For Likelihood analYsis;][]{2017Wilkinson}; the code determines the stellar population properties from the fit of the observed SEDs of galaxies with a set of stellar population models synthesized from stellar libraries, taking into account the initial mass function (IMF) and isochrones. This allows us to explore the formation and evolution of galaxies in the local universe as an archaeology approach \citep{2005Thomas}. The FIREFLY VAGC provides two different estimates of physical parameters with two different SED model templates (i.e. M11-MILES and MaStar). We present only the results based on the M11-MILES model templates for simplicity \citep{2011Maraston}, but note that our main conclusions do not change much even if we use the other estimates. 

\subsection{Stellar Age} \label{chap3.1} 

Figure \ref{Fig3} shows the mass fractions of different stellar populations for the four types of 191 SF-ETGs, 2541 q-ETGs, 2924 SF-LTGs, and 768 q-LTGs derived from the MaNGA FIREFLY VAC that provides the stellar age and mass per Voronoi bin \citep{2003Cappellari, 2022Neumann}. For a given type of galaxy, we determine the mass fraction for each age by combining the stellar mass from each Voronoi bin with the normalization by the total mass within the MaNGA field of view. We display the results derived from two different methods in determining the contributions of stellar populations to the overall galaxy spectral fit \citep{2017Wilkinson}. The mass-weighted age (left panel) better reflects the underlying stellar populations, while light-weighted age (right panel) is sensitive to recent star formation \citep{2017Wilkinson, 2022Xu}. 

First, we find that q-ETGs galaxies (gray line) show only a single peak with most of their stars forming simultaneously around 11-12 Gyrs ago (left panel). On the other hand, all of the other types have experienced more recent star formation. These galaxies show a clear second episode of star formation around 4-5 Gyrs ago, in addition to the first episode of star formation around 11-12 Gyrs ago. Interestingly, the age distribution of SF-ETGs (red line) is very different from that of q-ETGs, but is similar to that of SF-LTGs (blue line). SF-ETGs and SF-LTGs formed nearly half of their stars during the second star formation episode. These trends similarly appear even if we examine the samples in narrow mass ranges (see Figure \ref{Fig12} in Appendix).

 
\subsection{Radial Profile of Stellar Population} \label{chap3.2} 
\begin{figure*}[htb]
\centering
\includegraphics[bb = 30 530 320 780, width = 0.7 \linewidth, clip=]{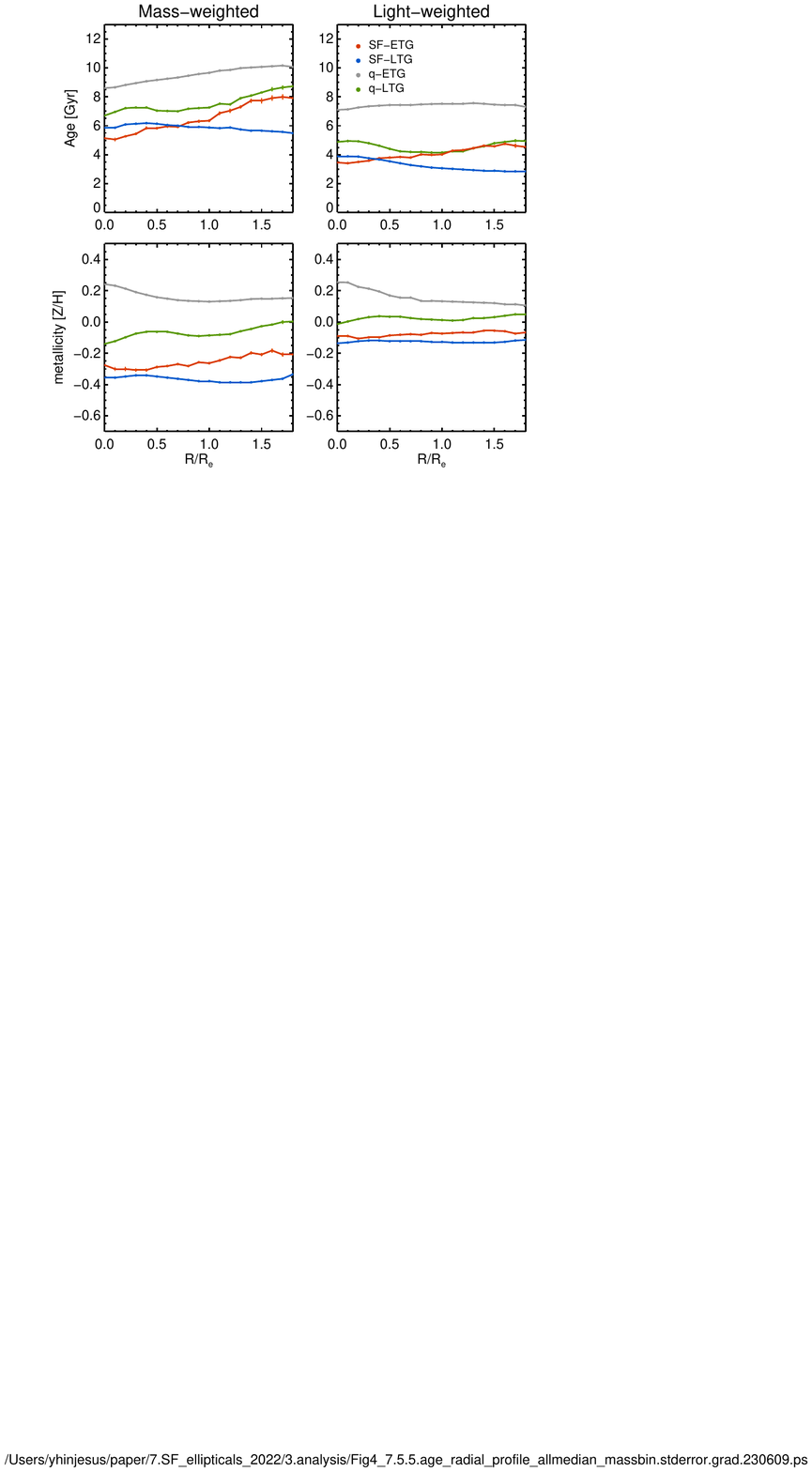}
\caption{Radial profiles of age (top) and stellar metallicity (bottom) for four different types of galaxies. The radial profiles are determined by calculating the median value at each binned radius after stacking the data for all of the galaxies in each type. The four types of galaxies are indicated by different colors. The age and metallicity estimates are influenced by whether they are weighted by mass (left) or light (right). The error bars are estimated by the bootstrap method. }\label{Fig4}
\end{figure*}


Figure \ref{Fig4} shows the radial profiles of stellar age (top panel) and metallicity (bottom panel) for the four types of galaxies. The gradients of stellar population can tell us whether they formed inside-out or outside-in, whether they have undergone mergers or gas accretion, and whether they have undergone quenching \citep{2019Miolino, 2019Taylor}. We take the median value of the stellar age or metallicity at each radial bin for each type of galaxy. Despite the large dispersion (i.e. the average scatters of age and metallicity for each sample are  $\sim$ 2.58 Gyrs and 0.25 dex, respectively.) the curves show different trends depending on galaxy type. 

The top panels of Figure \ref{Fig4} show the radial profiles of the mass- and light-weighted ages of the four types of galaxies. q-ETGs (gray line) are found to have positive and flat gradients in the mass- and light-weighted age, respectively. SF-ETGs (red line) also have positive gradients, which indicates younger populations in the central region (i.e. outside-in formation). In contrast, SF-LTGs (blue line) show negative gradients, which are in good agreement with the general feature of an old bulge and a younger disk for late-type galaxies \citep[i.e. inside-out galaxy formation;][]{1998Mo, 2012Pilkington, 2021Sharda}. It is interesting that SF-ETGs have an positive age gradient, which is very different from SF-LTGs even though they have similar star formation histories as seen in Figure \ref{Fig3}. 

Our results show that q-ETGs and SF-LTGs exhibit positive and negative age gradients, respectively. These findings are consistent with the results in previous studies of \citet{2017Goddard} and \citet{2021Parikh} who also studied early- and late-type galaxies separately. \citet{2019Lin} detected the negative age gradients for their sample of MaNGA galaxies (mostly late-type galaxies even though they did not separate the sample based on galaxy morphology as in our study), and found the importance of inside-out quenching along with other evidence. Similarly, \citet{2013Perez} used the galaxy sample of CALIFA survey (again they did not separate the sample based on galaxy morphology as in our study) and pointed out the importance of stellar mass in determining the age gradient (eventually inside-out and outside-in growths for more and less massive galaxies, respectively). We will discuss this radial dependence of stellar population regarding the formation of SF-ETGs in Section \ref{chap6}.



The bottom panels of Figure \ref{Fig4} show the radial profiles of the metallicity for the four types of galaxies. Quiescent galaxies (q-ETGs and q-LTGs) are found to be more metal-rich than star-forming galaxies (SF-ETGs and SF-LTGs), on average. The overall metallicity increases as galaxies have more stars formed at the first star formation episode shown in the left panel of Figure \ref{Fig3}. It is well consistent with the correlation between metallicity and age \citep{2000Kuntschner, 2005Thomas, 2005Gallazzi}. The metallicity has a correlation with galaxy stellar mass as well \citep{2000Kuntschner, 2005Thomas, 2005Gallazzi}. Therefore, it is expected that more massive galaxies would be more metal-rich and exhibit older stellar ages.

However, it is also possible for the metallicity of a galaxy to be lower because of the inflow of metal-poor gas from galaxy mergers or the cosmic web \citep{2010Rupke, 2016Ceverino}. Indeed, the lower metallicity observed in early-type galaxies compared to normal early-type galaxies is used as evidence to suggest that the recent star formation in those galaxies originates externally \citep{2022Jeong}. We will discuss it more in Section \ref{chap6}.

In terms of metallicity gradients, q-ETGs are found to have a negative gradient, but others appear nearly flat. We will discuss these gradients in more detail along with their mass dependence in Section \ref{chap6.2}, as they may depend on the mass of the galaxy.


\section{Gas Properties}\label{chap4}
To understand the properties of star formation in SF-ETGs, we investigate gas properties of SF-ETGs in comparisons with those of other types. We first examine the global properties, and then the spatially resolved properties.

\subsection{Global Properties} \label{chap4.1} 
\begin{figure*}[htb]
\centering
\includegraphics[bb = 30 640 310 780, width = 0.7 \linewidth, clip=]{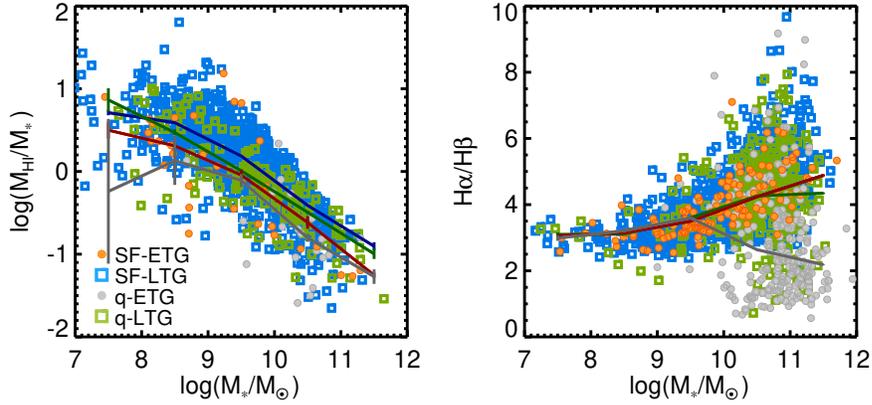}
\caption{Global gas and dust properties estimated by HI gas (left) and Balmer decrement (right) as a function of stellar mass. The four types of galaxies are represented by red (SF-ETGs), blue (SF-LTGs), gray (q-ETGs), and green (q-LTGs), respectively. We only display 10\% (left) and 1\% (right) of the data, respectively, for the clarity. The median values at each mass bin are displayed as solid lines with errors, which are estimated by the bootstrap method.  These quantities are calculated for the entire sample galaxies. }\label{Fig5}
\end{figure*}

Figure \ref{Fig5} shows the gas and dust properties of the four types of galaxies as a function of stellar mass. The left panel shows the atomic gas mass (HI) to stellar mass ratio. The atomic gas mass, $M_{\rm HI}$, is obtained from the Arecibo Legacy Fast ALFA Survey \citep[ALFALFA;][]{2005Giovanelli, 2011Haynes}. The plot shows that late-type galaxies (squares) have a higher gas fraction than early-type galaxies (circles) across the entire mass range, as expected. Also, if we focus on the galaxies with the same morphology, star-forming galaxies generally have more gas than quiescent galaxies.


The right panel shows the Balmer gradient (i.e. the ratio of the fluxes of the Balmer lines H$\alpha$ and H$\beta$), which can be a measure of dust extinction in galaxies; higher values of H$\alpha$/H$\beta$ indicate higher levels of dust extinction \citep{1936Berman, 2012Groves, 2013Hwang}. The H$\alpha$ and H$\beta$ fluxes are taken from the MPA/JHU DR7 VAGC \citep{2004Brinchmann}. In the absence of dust extinction, the H$\alpha$/H$\beta$ is expected to be 2.86 when assuming a temperature of $10,000 \rm K$ and an electron density of $100 \rm cm^{-3}$ \citep[case B in][]{2006Osterbrock}. The right panel of Figure \ref{Fig5} shows that massive q-ETG galaxies (gray line) have H$\alpha$/H$\beta$ values lower than 2.86, indicating little or no dust. However, SF-ETGs (red line) have H$\alpha$/H$\beta$ values similar to those of late-type galaxies, indicating a high level of dust extinction as in late-type galaxies.



\subsection{Spatial Distribution} \label{chap4.2} 

\begin{figure*}[htb]
\centering
\includegraphics[bb = 290 505 560 780, width = 0.6 \linewidth, clip=]{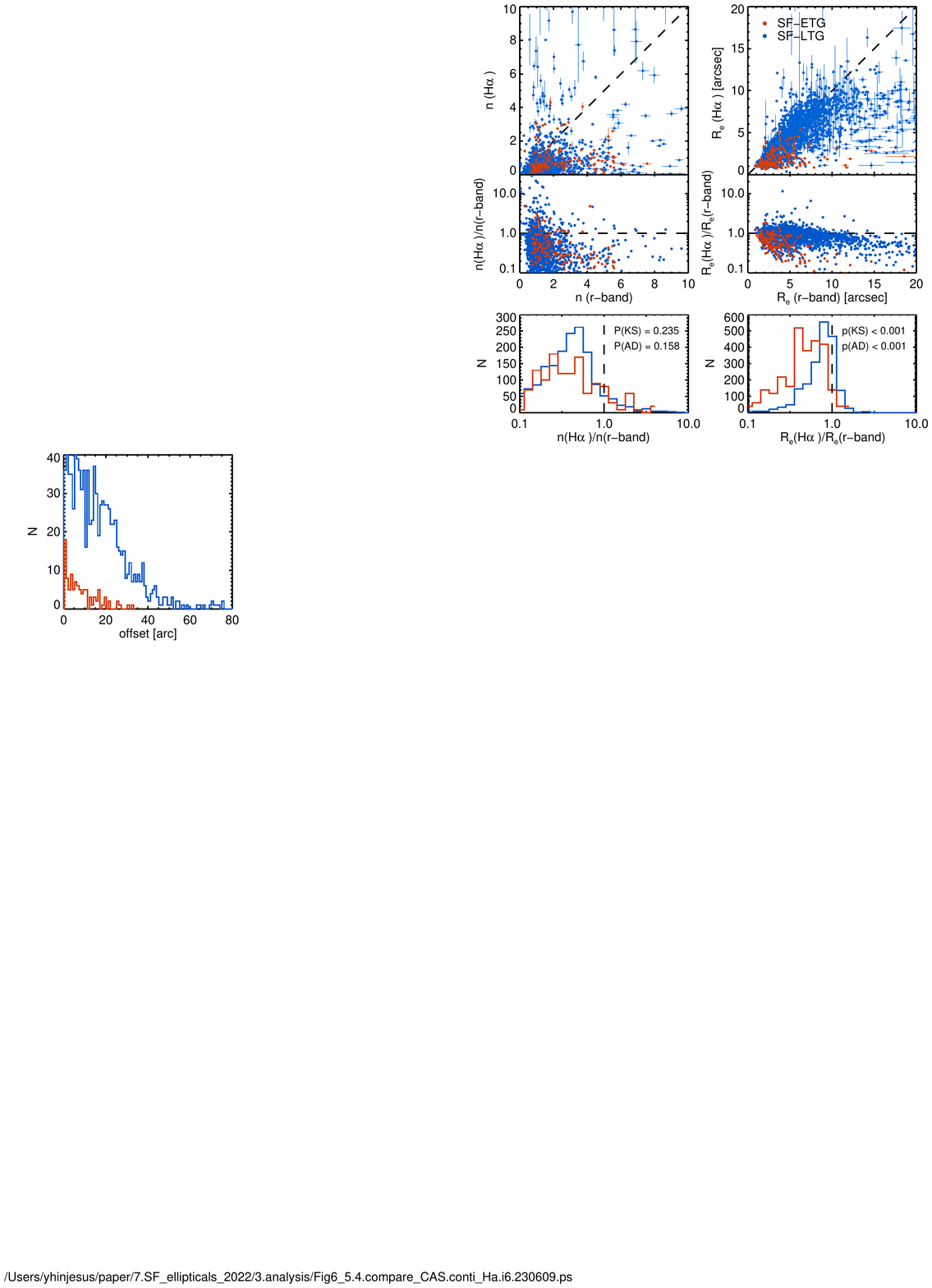}
\caption{Comparison of the sersic index (left) and effective radius (right) for the distribution between stars and gas measured from $r-$band and H$\alpha$ images, respectively. The ratios of these measurements between stars and gas are plotted as scatter diagrams (middle) and histograms (bottom). SF-ETGs and SF-LTGs are represented in red and blue. The histograms for SF-ETGs are multiplied by a factor of 10 on the left and of 20 on the right for the easy comparison with the SF-LTGs distribution. The {\it p}-values from the K-S and A-D tests for the ratio between SF-ETGs and SF-LTGs are shown in the top right of the bottom panels. The dashed lines on all panels indicate a one-to-one relation.}\label{Fig6}
\end{figure*}

To compare the spatial distribution of stars and gas in galaxies, we use the Sersic index ($n$) and the effective radius ($R_e$) measured from $r$-band (i.e. star) and H$\alpha$ (i.e. gas) images of SDSS/MaNGA. To do that, we apply the GALFIT software \citep{2002Peng} to these images. Because the area coverage of H$\alpha$ images is often smaller that that of $r$-band images, we run the GALFIT for the region common between H$\alpha$ and $r$-band images for each galaxy. As a result, we obtain $n$ and $R_e$ values for 185 out of 191 SF-ETGs and 2226 out of 2924 SF-LTGs. The top panels of Figure \ref{Fig6} display $n$ (left panel) and $R_e$ (right panel) measurements from $r$-band and H$\alpha$ images for both SF-ETGs (red) and SF-LTGs (blue). To better show the difference of the distribution between stars and gas, we also plot the ratios of the measurements between stars and gas as scatter diagrams in the middle panels and histograms in the bottom panels. 

The left panels of Figure \ref{Fig6} show that the Sersic index ($n$) of the gas is generally lower than that of the stars for both SF-ETGs and SF-LTGs; the histogram for the ratios in the bottom panel is skewed toward lower values. In particular, most SF-ETGs have $n$ values smaller than 2 for the gas, which indicates that the gas is distributed exponentially as for stellar disks. 

The right panels of Figure \ref{Fig6} show that there is a difference in the distribution of effective radius ($R_e$) between SF-ETGs and SF-LTGs; the effective radius of the H$\alpha$ distribution of SF-ETGs is generally smaller than that of stellar continuum distribution, which is not true for SF-LTGs (i.e. similar distributions between H$\alpha$ and stellar continuum). The ratio of the distribution between stars and gas is peaked around one for SF-LTGs, but is skewed toward lower values for SF-ETGs (bottom panel). This suggests that in SF-ETGs gas is more centrally concentrated compared to the stellar distribution. Centrally concentrated gas has been observed in blue elliptical galaxies \citep{2008Lee} as well, along with the misalignment between stars and gas, which is independent of galaxy morphology \citep{2016Chen, 2016Jin, 2022Xu}. We run the Kolmogorov-Smirnov (K-S) and Anderson-Darling (A-D) tests to determine whether SF-ETGs and SF-LTGs are drawn from the same distribution. In the bottom panels, we provide the corresponding {\it p}-value, which represents the probability that two samples are drawn from the same parent distribution. The {\it p}-value demonstrates that the distribution of gas in SF-ETGs differs from that in SF-LTGs, particularly in relation to $R_e$.



\subsection{Kinematics of Gas} \label{chap4.3} 

\begin{figure*}[htb]
\centering
\includegraphics[bb = 20 280 420 420, width = 0.95 \linewidth, clip=]{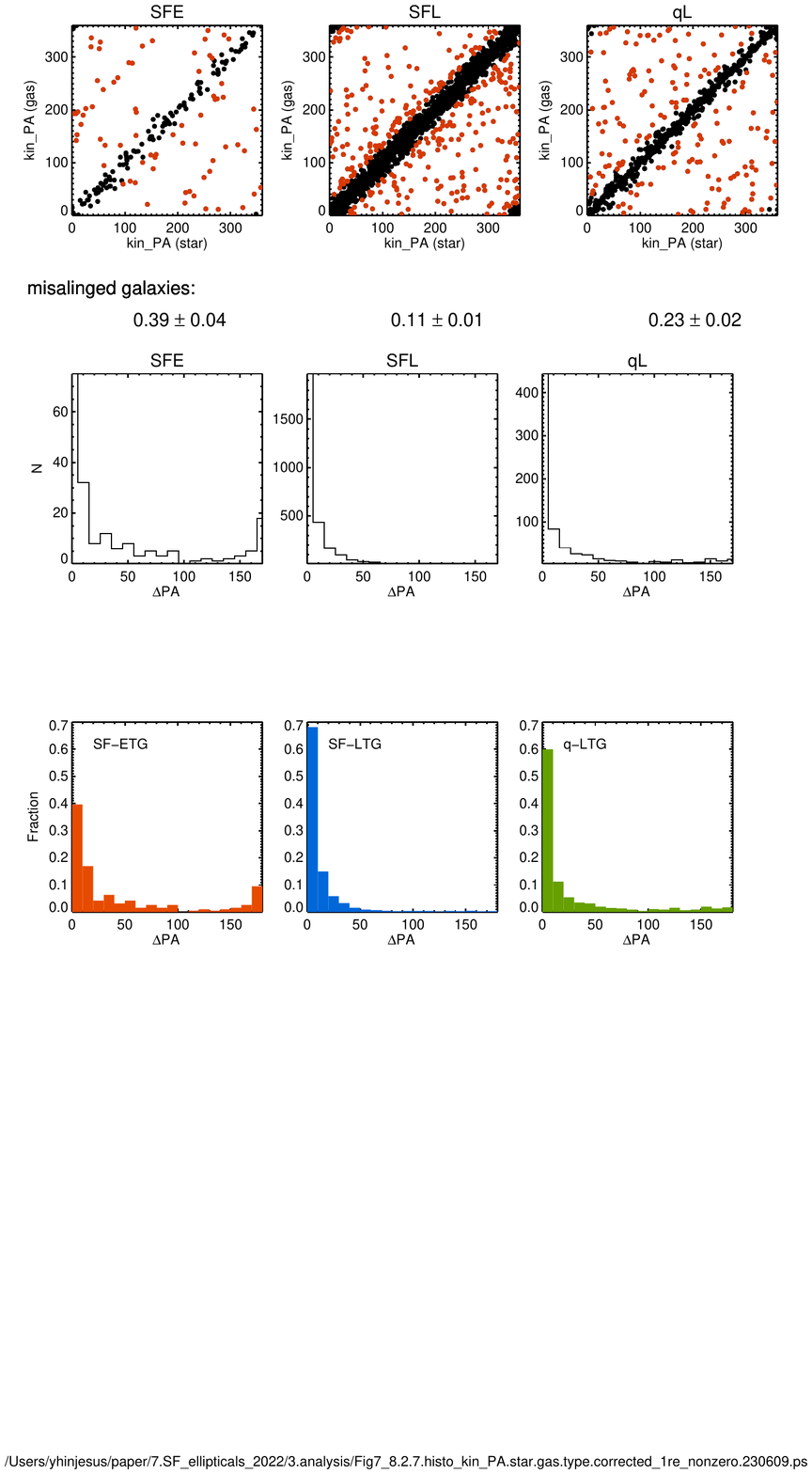}
\caption{Fractions of position angle (PA) offsets of the rotation axes between stars and gas for three different types of galaxies: SF-ETGs (left), SF-LTGs (middle), and q-LTGs (right). Galaxies with $\Delta\rm PA > 30^{\circ}$ are classified as misalignments, and galaxies with $\Delta\rm PA > 150^{\circ}$ are selected as counter-rotators. }\label{Fig7}
\end{figure*}

To further understand the kinematic difference between stars and gas, we measure the kinematic position angle (PA) of the stellar and gas rotation axes by applying the \textsc{Fit kinematic pa} routine \citep{2007Cappellari, 2011Krajnovi} to velocity maps of stars and gas (i.e. H$\alpha$) from the MaNGA survey. The routine generates a model map for each possible PA value using the observed velocity map, and determines the kinematic PA that minimizes the $\chi^2$ between observed and model maps. We utilize the velocity maps within 1$R_e$ to avoid large errors from the outer region with low S/N ratio \citep{2019Bryant}. We could obtain the kinematic PAs 
for both stars and gas for 189, 2888, and 740 objects out of 191 SF-ETGs, 2924 SF-LTGs, and 768 q-LTGs, respectively. Because the \textsc{Fit kinematic pa} routine does not consider the direction of rotation, we redefine the PAs to the counter-clockwise angle from north to the receding side of the velocity map to distinguish counter-rotators. 

Figure \ref{Fig7} displays the PA offsets ($\Delta\rm PA$) of the rotation axes between stars and gas for the three types of galaxies. The middle panel shows that most SFLs (89$\%$) have aligned gas with stars within 30$^\circ$. However, SF-ETGs (left panel) and q-LTGs (right panel) have a wide range of $\Delta$PA. In Table \ref{Table1}, we define misaligned galaxies and counter-rotators as those with $\Delta$PA values larger than $30^{\circ}$ and $150^{\circ}$, respectively \citep{2011Davis, 2016Chen, 2019Bryant}. According to this definition, counter-rotators are also selected as misaligned galaxies. The misaligned galaxies are common among SF-ETGs ($39\%$). The fraction of counter-rotators among SF-ETGs is relatively high ($14\%$) compared to other types. We note that q-LTGs have a relatively higher fraction of misalignments and counter-rotators than SF-LTGs. The misalignment between stars and gas may indicate an external origin of the gas through the mechanisms including gas accretion from the cosmic web, galaxy interactions, or mergers \citep{1994Rubin, 1997Thakar, 2016Chen, 2016Jin}. We will discuss it in Section \ref{chap6.1}.

\begin{table}[htb]
  \caption{Fractions of misaligned galaxies and counter-rotators among SF-ETGs, SF-LTGs, and q-LTGs. According to the definition used, counter-rotators are also selected as misaligned galaxies. The uncertainties indicate the confidence interval of the fraction for a beta distribution, which depends on the sample size \citep{2011Cameron}.} \label{Table1}
 \begin{center}
 \begin{tabular}{llllll}
 \hline
 \hline
  \multicolumn{2}{c}{Classification} &
  \multicolumn{2}{c}{Misalignment} &
  \multicolumn{2}{c}{Counter-rotator} \\
  \multicolumn{2}{c}{} &
  \multicolumn{2}{c}{($\Delta\rm PA > 30^{\circ}$)} &
  \multicolumn{2}{c}{($\Delta\rm PA > 150^{\circ}$)} \\
 \hline
SF-ETGs & (189) & $39\%\pm0.04$ & (73) & $14\%\pm0.03$ & (26) \\
SF-LTGs & (2888) & $11\%\pm0.01$ & (307) & $1\%\pm0.002$ &(34) \\ 
q-LTGs &(740) & $23\%\pm0.02$ & (173) & $5\%\pm0.01$ & (38) \\ 
 \hline
 \multicolumn{6}{l}{\footnotesize The numbers in the parentheses indicate the    number of galaxies.}\\
 \end{tabular}
 \end{center}
 \end{table}

\section{Environments} \label{chap5}

\begin{figure*}[htb]
\centering
\includegraphics[bb = 30 500 300 750, width = 0.75 \linewidth, clip=]{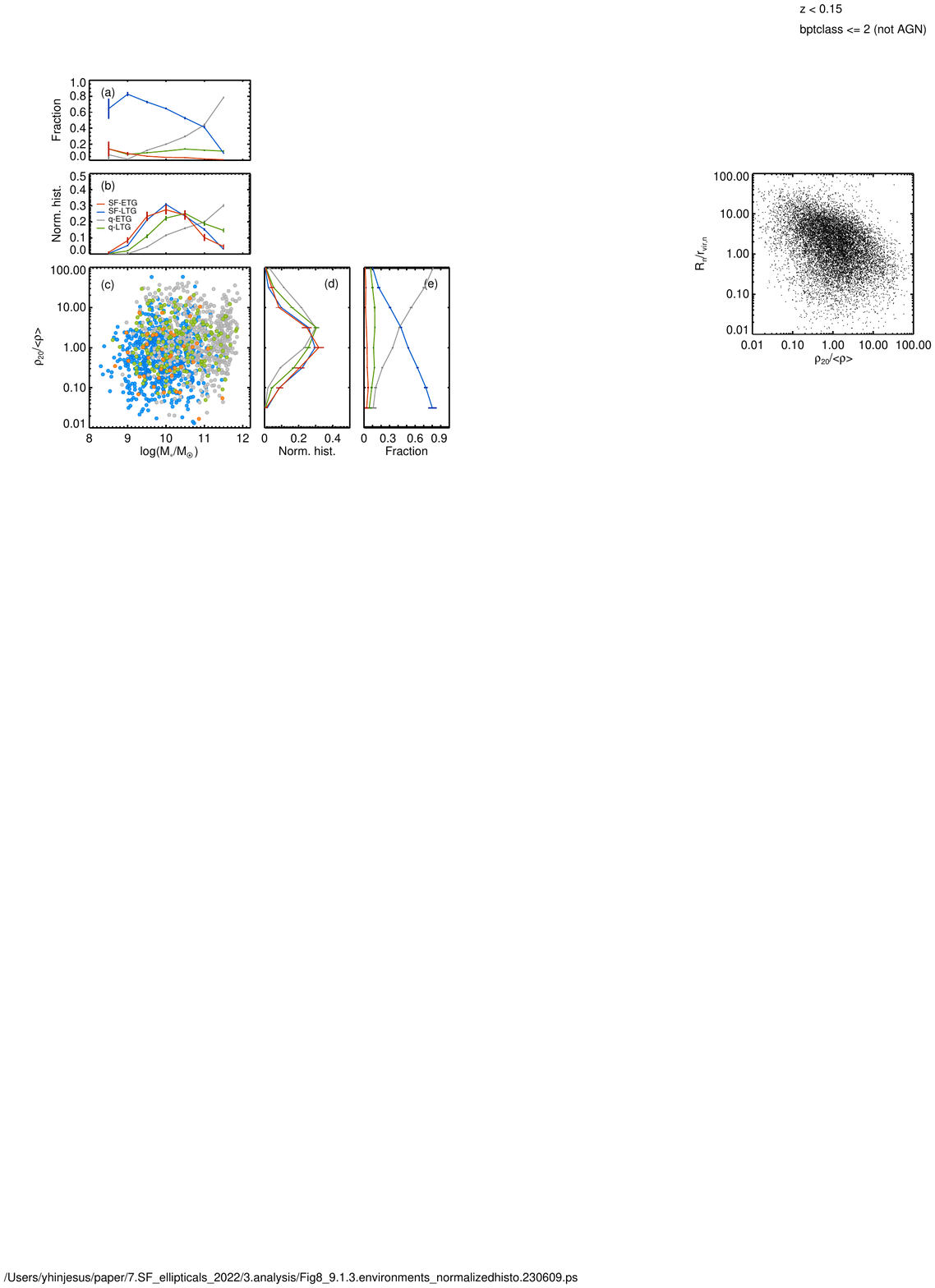}
\caption{Distributions of galaxies on the plane of the stellar mass $\rm log(M_{\ast}/M_{\odot})$ and the environmental parameter $\rho_{20}/<\rho>$ (c), with 20\% of galaxies for each type represented by different colors: SF-ETGs (red), SF-LTGs (blue), q-ETGs (gray), and q-LTGs (green). The fractions ((a) and(e)) and the normalized histograms ((b) and (d)) are plotted as functions both of the stellar mass (upper panels) and of the environmental parameter (right panels). These quantities are calculated for the entire sample galaxies. The uncertainties are estimated from the confidence interval of the fraction for a beta distribution. }\label{Fig8}
\end{figure*}

\begin{figure}[htb]
\centering
\includegraphics[bb = 40 340 190 530, width = 0.95 \linewidth, clip=]{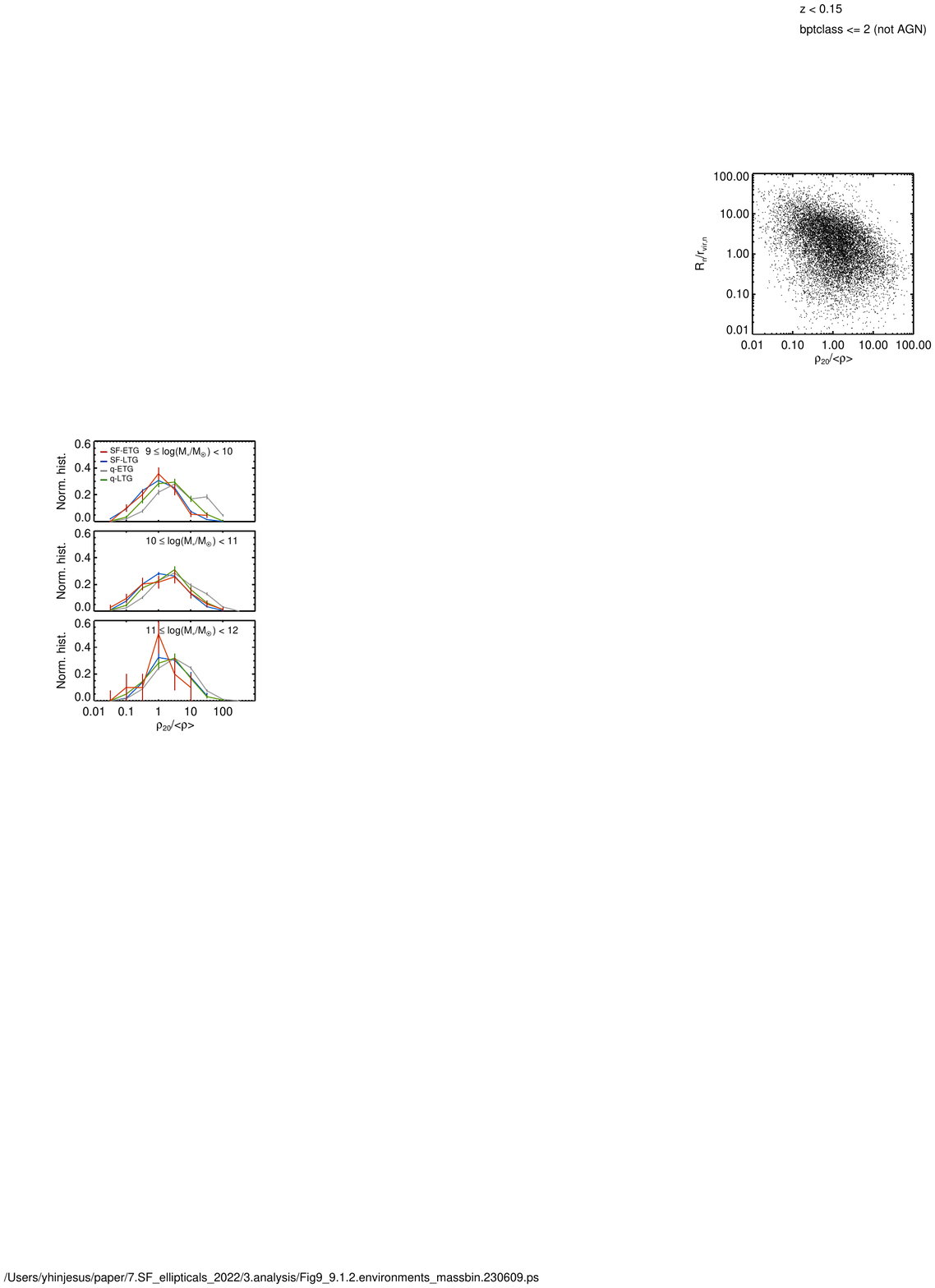}
\caption{Normalized histograms of four types of galaxies as a function of the large-scale environment $\rho_{20}/<\rho>$, with the three panels representing galaxies in different mass ranges from lower-mass (top) to higher-mass (bottom). The four types are represented by different colors, SF-ETGs (red), SF-LTGs (blue), q-ETGs (gray), and q-LTGs (green). The uncertainties are estimated using the confidence interval of the fraction for a beta distribution. } \label{Fig9}
\end{figure}

To study the environment of galaxies, we use a background mass density ($\rho_{20}$) at a given location of a galaxy. This measure is based on the $r-$band luminosity $L$ of the 20 closest galaxies and is used to quantify the large-scale environment \citep{2008Park, 2009Park}. It is calculated using the following equation: 
\begin{eqnarray}\label{eq2}
\rho_{20}(\textbf{x})/<\rho>=\sum_{i=1}^{20}\gamma_{i}L_{i}W_{i}(|\textbf{x}_{i}-\textbf{x}|)/<\rho> 
\end{eqnarray}
where $<\rho>$ is the mean density of the universe, $\gamma$ is the mass-to-light ratio of a galaxy, and $W(x)$ is a smoothing filter function. The mean mass density within a sample of total volume $V$ can be calculated by 
\begin{eqnarray}\label{eq3}
<\rho>=\sum_{\rm all}\gamma_{i}L_{i}/V
\end{eqnarray}
where the summation is for galaxies brighter than $M_{r}=-19.0$ in the sample. The details for this calculation of $\rho_{20}/<\rho>$ are given in \citet{2018Lee} where we adopt the value of each galaxy in this study.

Figure \ref{Fig8} displays the distribution of galaxies on the plane of the stellar mass $\rm log(M_{\ast}/M_{\odot})$ and the environmental parameter $\rho_{20}/<\rho>$ (Figure \ref{Fig8}(c)). The upper panels show the fraction and the normalized histogram of the stellar mass, while the right panels show such plots for the environmental parameter. 

The top panel of Figure \ref{Fig8} shows the fractions of four types of galaxies in different mass bins. This shows that most low-mass galaxies are SF-LTGs (blue line), while q-ETGs (gray line) become abundant among more massive galaxies. SF-ETGs (red line) and q-LTGs (green line) make up a relatively small fraction of the total. However, the normalized histogram in Figure \ref{Fig8}(b) provides a different perspective on the distribution of each type of galaxy in the mass range. The histogram shows the number of galaxies for each type at a given mass bin, normalized by the total number of galaxies for each type. This shows that SF-ETGs and SF-LTGs are more common in low-mass galaxies around $10^{10}M_{\odot}$, while q-LTGs and q-ETGs are prevalent among more massive galaxies, in that order. This is consistent with the downsizing of galaxies where the quenching of star-formation occurs first in more massive galaxies and then in less-massive ones \citep{1996Cowie, 2007Noeske, 2011Brammer}.

The right panel of Figure \ref{Fig8} shows the fractions of four types of galaxies in different environmental parameter, $\rho_{20}/<\rho>$. This shows that q-ETGs are more common in higher-density environments, while SF-LTGs are more prevalent in lower-density environments. This is consistent with the morphology-density relation, which is that the fraction of elliptical galaxies increases with local density \citep{1980Dressler, 1984Postman}. SF-ETGs and q-LTGs are relatively rare in all environments. However, the normalized histogram in the bottom middle panel of Figure \ref{Fig8} shows that SF-ETGs are peaked at lower-density environments together with SF-LTGs, while q-LTGs are skewed towards higher-density environments, similar to q-ETGs. Overall, these results suggest that the star-formation is more likely to occur in less-massive galaxies (i.e. internal effect) and in lower-density environments (i.e. external effect) regardless of their morphology (early- and late-type). This is consistent with the results in previous studies; the quenching of star formation is more effective in more massive galaxies and in denser environments \citep{2006Bundy, 2008Cooper, 2009Tasca, 2010Peng}.

To further examine the environmental difference among different galaxy types without mass effect, we show the normalized histograms of each type of galaxy as a function of $\rho_{20}/<\rho>$ divided into three mass bins. These histograms show that star-forming galaxies tend to be found in lower-density environments, regardless of their morphology. Table \ref{Table2} shows the {\it p}-value from the K-S and A-D tests of $\rho_{20}/<\rho>$ distributions among different types of galaxies at each mass bin. The large {\it p-}values for the tests between SF-ETGs and SF-LTGs indicate that they are not significantly different from each other in terms of their environments across the entire mass range. On the other hand, the other types show small {\it p}-values, indicating that they tend to reside in different environments. However, in the highest mass bin, the effect of environment is weaker than for lower mass bins, which is consistent with the results in previous studies in sense that galaxies above $10^{10.6} \rm M_\odot$ are more influenced by their intrinsic nature (i.e., their mass) than by their environments \citep{2009Tasca, 2010Peng}.

\begin{deluxetable*}{ccccc}
 \tablecaption{{\it P}-values from the K-S (upper rows) and A-D (lower rows) tests for the background mass density $\rho_{20}/<\rho>$ between different galaxy types at given mass bins. \label{Table2}}
 \tablehead{
 \colhead{mass range} &
 \colhead{SF-ETG versus q-ETG} &
 \colhead{SF-LTG versus q-LTG} &
 \colhead{SF-ETG versus SF-LTG} &
 \colhead{q-ETG versus q-LTG} \\
}
\startdata
$9\leq \rm log(M_\star/M_\odot) < 10$ & $< 0.001$ & $< 0.001$ & 0.678 & $< 0.001$ \\
                                       & $< 0.001$ & $< 0.001$ & 0.570 & $< 0.001$ \\
$10\leq \rm log(M_\star/M_\odot) < 11$ & 0.001 & $< 0.001$ & 0.816 & $< 0.001$ \\ 
                                       & $< 0.001$ & $< 0.001$ & 0.664 & $< 0.001$ \\ 
$11\leq \rm log(M_\star/M_\odot) < 12$ & 0.063 & 0.983 & 0.509 & $< 0.001$ \\ 
                                       & 0.051 & 0.955 & 0.360 & $< 0.001$ \\ 
\enddata
\end{deluxetable*}

\section{Discussion}\label{chap6}
\subsection{What Makes the Star Formation Activity in SF-ETGs?} \label{chap6.1}

In this study, we have examined the properties of SF-ETGs and have compared them to those of other types. We find that SF-ETGs are distinct from q-ETGs but show some properties similar to SF-LTGs including stellar populations, gas and dust content, stellar mass, and environments. SF-ETGs show two distinctive star formation episodes; this is similar to both late-types (i.e. SF-LTGs and q-LTGs), but is different from q-ETGs that have a single star formation episode occurred around 11-12 Gyrs ago (Figure \ref{Fig3}). In addition, SF-ETGs have younger stellar populations, lower metallicities (Figure \ref{Fig4}), and more gas/dust (Figure \ref{Fig5}) than q-ETGs.

SF-ETGs and SF-LTGs show similar distributions in terms of stellar mass and environment in lower-mass and lower-density environments, despite different morphologies (Figure \ref{Fig8}). This is consistent with the observed star formation rate (SFR)-density relation in the local universe \citep{1998Balogh, 2002Lewis, 2003Gomez, 2004Balogh, 2005Christlein, 2010Hwang, 2019Hwang}, regardless of their morphology. The SFR tends to be lower in the high-density environments due to various mechanisms; these including ram-pressure stripping and tidal disruption that can inhibit star formation in dense environments \citep{2000Quillis}, and frequent galaxy interactions in dense regions that can deplete the gas supply and create a gas-poor environment \citep{2018Pan, 2009ParkHwang, 2019Hwang}. As a result, the lower-density environments are more favorable for star formation in the local universe; this applies to the current star formation in SF-ETGs and SF-LTGs.

However, SF-ETGs and SF-LTGs are differentiated by their gas distributions and kinematics. The gas in SF-ETGs is often more centrally concentrated (Figure \ref{Fig6}) and may rotate misaligned with the stellar rotation (Figure \ref{Fig7}), while the gas in SF-LTGs tends to be spread out across the disk of galaxy (Figure \ref{Fig6}) and rotate in a manner similar to the stars (Figure \ref{Fig7}). The alignment between stellar and gas rotations in a galaxy can provide information on the origin of the gas. If gas and stars are well-aligned, it is likely that the gas has an internal origin, such as gas produced through the stellar evolution of the galaxy \citep{2006Sarzi, 2016Chen, 2016Jin}. On the other hand, the misalignment between stars and gas can indicate an external origin for the gas, such as gas accretion, galaxy interactions, or mergers \citep{1994Rubin, 1997Thakar, 2016Chen, 2016Jin}. 

Observations have shown that the kinematic PAs of stars and gas are usually well-aligned in non-interacting galaxies \citep{2014Barrera-Ballesteros}, but are often misaligned in interacting galaxies \citep{2015Barrera-Ballesteros}. Numerical simulations have demonstrated that major mergers of gas-rich disks can produce misalignments or counter-rotating gas, while gas-poor major mergers tend to produce counter-rotating systems \citep{1991Hernquist, 2007Jesseit, 2007Naab}. In cosmological simulations, counter-rotating systems are also found, caused by minor mergers and the accretion of gas from filaments \citep{2015Taylor, 2018Taylor, 2019Taylor}. These could be understood with the idea that the accreted gas exchanges angular momentum with pre-existing gas, falling to the center of the galaxy. Therefore, external origins, such as merger or accretion, could be responsible for the star formation of SF-ETGs that show higher fractions of misalignments or counter-rotating gas.

\subsection{Possible Mechanisms of the Formation for SF-ETGs} \label{chap6.2}

\begin{figure*}[htb]
\centering
\includegraphics[bb = 0 530 450 800, width = 0.9 \linewidth, clip=]{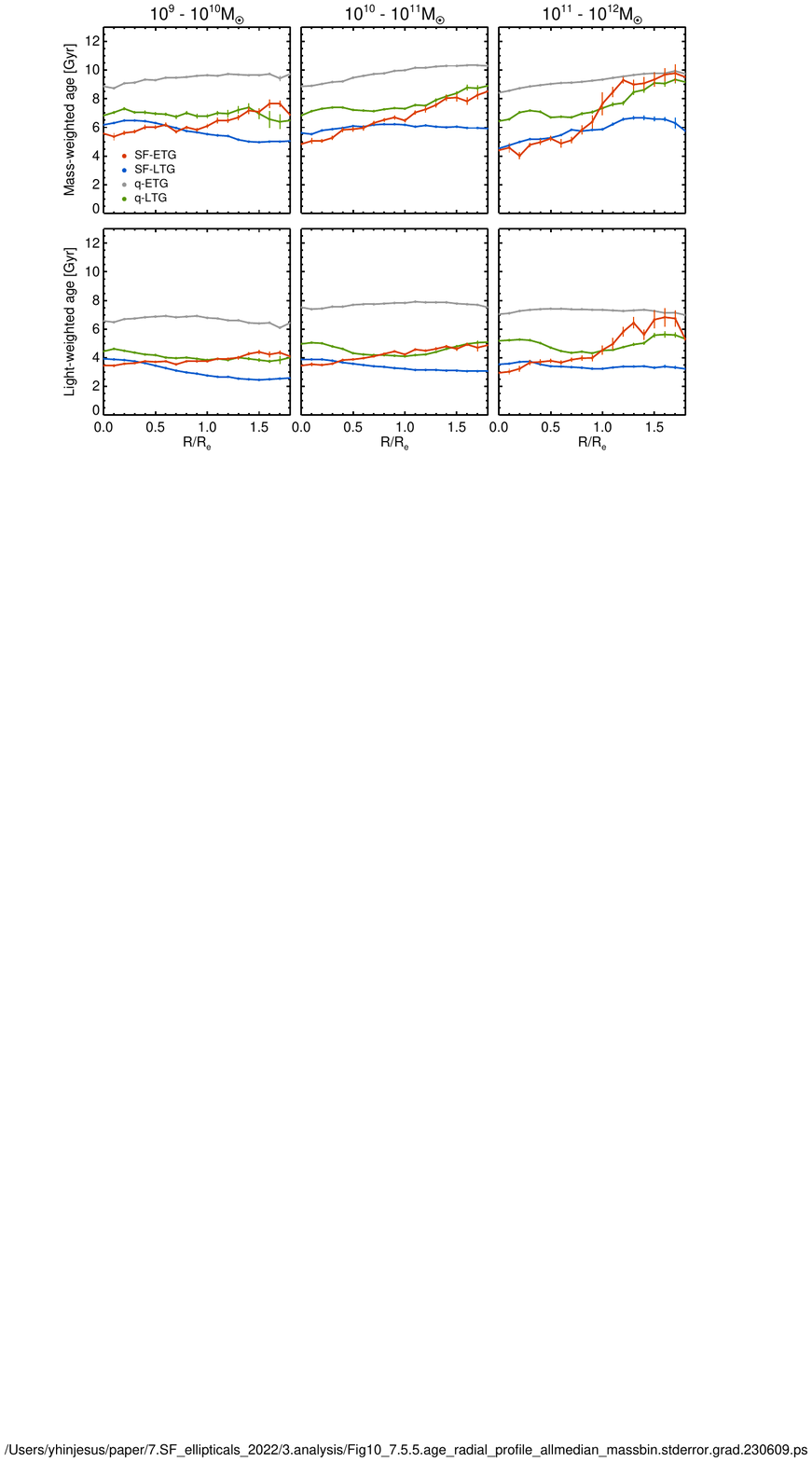}
\caption{Radial profiles of stellar age for four types of galaxies in different mass bins. These profiles are measured in the same way as in Figure \ref{Fig4}, but are displayed by the three mass bins. The different colors represent different types.  The age is estimated by mass-weighted (top) and light-weighted (bottom) ways, respectively. The errors are measured by the bootstrap method. }\label{Fig10}
\end{figure*}

\begin{figure*}[htb]
\centering
\includegraphics[bb = 0 530 450 800, width = 0.9 \linewidth, clip=]{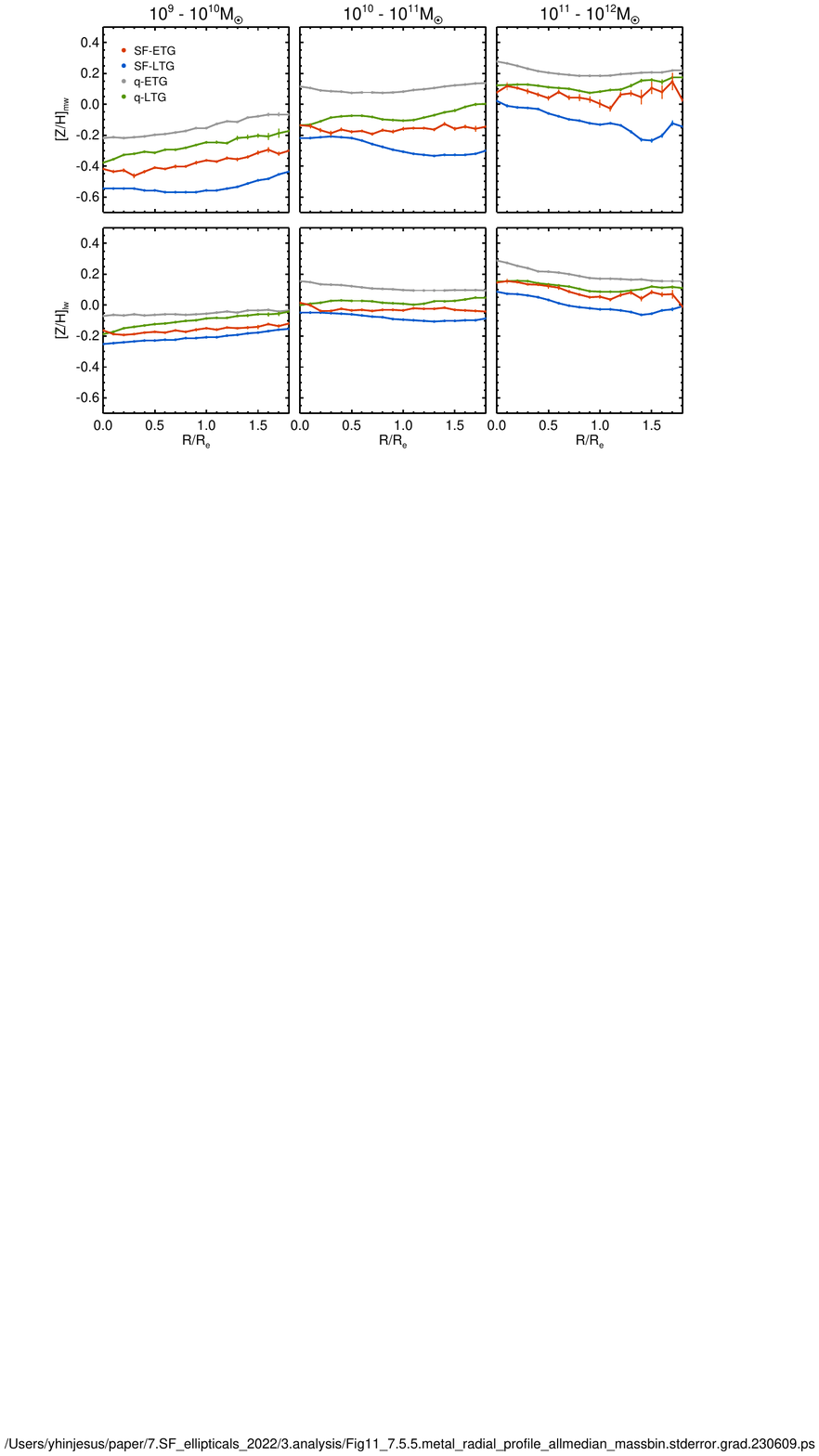}
\caption{Radial profiles of stellar metallicity for four types of galaxies in different mass bins. These profiles are measured in the same way as in Figure \ref{Fig3}, but are displayed by the three mass bins. The metallicity is estimated by mass-weighted (top) and light-weighted (bottom) ways, respectively. The errors are calculated by the bootstrap method.}\label{Fig11}
\end{figure*}


To further examine the age gradients for the four types of galaxies, we show them in Figure \ref{Fig10} that is similar to Figure \ref{Fig4} but divided by three mass bins. This figure shows that SF-ETGs generally have positive age gradient, while SF-LTGs tend to have negative gradients. This indicates that SF-ETGs have a younger population in their central region, whereas SF-LTGs have a younger population in the outer region. These trends appear consistent with the result for the gas distribution in Figure \ref{Fig6}, which shows centrally concentrated gas in SF-ETGs and widely spread gas in SF-LTGs.

Our results could be understood with the help of cosmological simulations. For example, \citet{2011Wang} used the Acquaris simulation \citep{2008Springel}, and showed that different mechanisms have different effects on the internal structure of galaxies. Major merger can drive gas to the central region of a galaxy, which is efficient when the progenitor carrying the gas is massive. On the other hand, the gas from minor mergers or from filaments generally ends up being in the outer regions of galaxies \citep{2011Wang, 2012Lackner, 2012Moran}. The mass of galaxy is also an important factor in this process. The accreted gas tends to sink towards the center of galaxies in less massive galaxies \citep{2006Dekel, 2014Sanchez}, but primarily makes up the outer regions in massive galaxies \citep{2011Wang, 2012Lackner, 2014Sanchez}. As a result, it is plausible that major mergers play an role in the formation of massive SF-ETGs, while the accretion gas from the cosmic web could account for less-massive SF-ETGs. We note that the age gradient of SF-ETGs becomes more pronounced as the mass increases, and SF-LTGs also show a positive age gradient in the highest mass bin (Figure \ref{Fig10}).

Figure \ref{Fig11} displays the metallicity gradients for the four types of galaxies, divided by three mass bins. We find that the mass of galaxy is more influential on the metallicity gradient than the galaxy type. Galaxies in the highest mass bin tend to have negative metallicity gradients, while those in the lowest mass bin tend to have positive gradients. 

Many Observations and simulations have shown that most galaxies tend to have negative metallicity gradients \citep{1995Edmunds, 1998VanZee, 1999Molla, 2010Kewley, 2017Belfiore, 2021Hemler, 2022Porter}, which is consistent with the inside-out galaxy formation \citep{1998Mo, 2012Pilkington, 2021Sharda}. Early star formation rapidly increases the metallicity in the centeral region of a galaxy, while the outer disk increases in metallicity more slowly. As a result, the metallicity gradient becomes shallower over time \citep{1997Molla, 2007Magrini}. 

On the other hand, recent mergers or interactions can lead to flattened metallicity gradients and the lower central metallicities in galaxies \citep{2008Rupke, 2008Ellison, 2008Michel-Dansac, 2010Kewley}. In a merger scenario, galaxy interactions can drive large amount of gas into the central regions of a galaxy, mixing the gas and diluting its central metallicity \citep{2010Kewley}. Simulation studies have shown that the metallicity gradient becomes significantly flatter and the central metallicity is diluted after the first pericenter passage in a major merger event \citep{2010Rupkea, 2010Montuori}. These galaxy mergers and interactions could explain why there is little metallicity gradient in SF-ETGs compared to that in SF-LTGs at $M_{\ast} > 10^{10} M_{\odot}$.

Lastly, we note that misalignments or counter-rotating gas can contribute to the bulge formation or to the morphological transformation \citep{2009Scannapieco, 2011Pichon, 2012Sales, 2015Zolotov, 2019Park}. Simulations have shown that misalignments are a common occurrence and lead to the redistribution of mass from the disk to the spheroid \citep{2009Scannapieco, 2012Sales}. Observations show the results consistent with those in simulations in the sense that counter-rotators are predominantly present in early-type galaxies than late-type galaxies \citep{2001Kannappan}, and are found more frequently in rounder and slow-rotating galaxies than flatter and fast-rotating galaxies \citep{2006Sarzi, 2011Davis}. Therefore, we suggest that the progenitors of SF-ETGs could be SF-LTGs that have similar properties and reside in similar environments. Some of SF-LTGs might experience morphological transformation into SF-ETGs through recurrent infalling of misaligned gas from galaxy mergers or from the cosmic web. 

\section{Conclusions} \label{chap7}
To better understand the origin of current star formation in SF-ETGs, we have used a sample of $\sim$580,000 galaxies at z $<$ 0.15 from SDSS/DR17. Galaxies were classified into four types based on their morphology and the star formation rate (SFR): star-forming early- (SF-ETG) and late-type (SF-LTH) galaxies, and quiescent early- (q-ETG) and late-type (q-LTG) galaxies. We examined galaxy properties of these four types of galaxies including their stellar populations, gas and dust content, stellar mass, and environments. In particular, spatially resolved properties of the stellar age, metallicity, velocity, and H$\alpha$ emission were investigated for 10,064 galaxies that have IFS data from MaNGA/DR17. 

Our main results are as follows.
\begin{enumerate}
\item The analysis of the optical spectra shows that SF-ETGs have two distinctive episodes of star formation, which is similar to late-type galaxies. In contrast, q-ETGs have only one episode of star formation that occurred about 11-12 Gyrs ago. 

\item The SF-ETGs have some properties similar to those of SF-LTGs, which include a larger amount of gas and dust, younger populations, and lower metallicity than q-ETGs. Both SF-ETGs and SF-LTGs are found in less massive systems and in lower-density environments than q-ETGs.

\item While SF-ETGs and SF-LTGs share many properties, they are distinguished by their spatially resolved characteristics such as the distribution of the stellar populations and gas. SF-ETGs tend to have more concentrated gas in their central regions and often have kinematically misaligned gas and stars. They also tend to have a positive age gradient, indicating the presence of younger populations in the central region. In contrast, SF-LTGs tend to have gas that is more evenly distributed throughout the disk and corotates with the stars. They tend to have a negative age gradient.
\end{enumerate}

Our observational data support that external factors such as galaxy mergers or accretion of gas from the cosmic web play a role in the current star formation in SF-ETGs. The SF-LTGs could be the progenitors of SF-ETGs as they share similar properties and reside in similar environments. The recurrent inflow of misaligned gas could drive the morphological transformation from SF-LTGs to SF-ETGs. In the future, we plan to use cosmological hydrodynamical simulations such as Horizon Run 5 \citep{2021Lee} and IllustrisTNG \citep{2019Nelson} to directly trace the formation and evolution of SF-ETGs.


\section*{Acknowledgement}
We express our gratitude to the anonymous referee for the valuable comments, which greatly contributed to the improvement of this paper. We would also like to extend our appreciation to Ena Choi for the insightful discussions. This research was supported by Basic Science Research Program through the
National Research Foundation of Korea (NRF) funded by the Ministry of Education (RS-2023-00249435). HSH acknowledges the support by the National Research Foundation of Korea (NRF) grant funded by the Korea government (MSIT) (No. 2021R1A2C1094577).

Funding for the Sloan Digital Sky Survey IV has been provided by
the Alfred P. Sloan Foundation, the U.S. Department of Energy Office
of Science and the Participating Institutions. SDSS-IV acknowledges
support and resources from the Center for High-Performance Computing at the University of Utah. The SDSS web site is www.sdss.org.

SDSS-IV is managed by the Astrophysical Research Consortium
for the Participating Institutions of the SDSS Collaboration including
the Brazilian Participation Group, the Carnegie Institution for Science, Carnegie Mellon University, the Chilean Participation Group,
the French Participation Group, Harvard-Smithsonian Center for Astrophysics, Instituto de Astrofísica de Canarias, The Johns Hopkins
University, Kavli Institute for the Physics and Mathematics of the
Universe (IPMU) / University of Tokyo, Lawrence Berkeley National
Laboratory, Leibniz Institut für Astrophysik Potsdam (AIP), MaxPlanck-Institut für Astronomie (MPIA Heidelberg), Max-PlanckInstitut für Astrophysik (MPA Garching), Max-Planck-Institut für
Extraterrestrische Physik (MPE), National Astronomical Observatories of China, New Mexico State University, New York University,
University of Notre Dame, Observatário Nacional / MCTI, The Ohio
State University, Pennsylvania State University, Shanghai Astronomical Observatory, United Kingdom Participation Group, Universidad
Nacional Autónoma de México, University of Arizona, University of
Colorado Boulder, University of Oxford, University of Portsmouth,
University of Utah, University of Virginia, University of Washington,
University of Wisconsin, Vanderbilt University and Yale University.

\bibliography{2022_SFE.bib} 
\begin{appendices}
\renewcommand\thefigure{A\arabic{figure}} 
\section*{Appendix: Stellar populations of galaxies with different mass} 
To investigate whether there is any difference in stellar population among different types of galaxies by fixing stellar mass ranges, we present the mass fractions of stellar populations, divided by three mass bins. The main features remain similar regardless of the stellar mass.
\end{appendices}


\begin{figure*}[htb]
\centering
\includegraphics[bb = 0 520 450 800, width = 0.9 \linewidth, clip=]{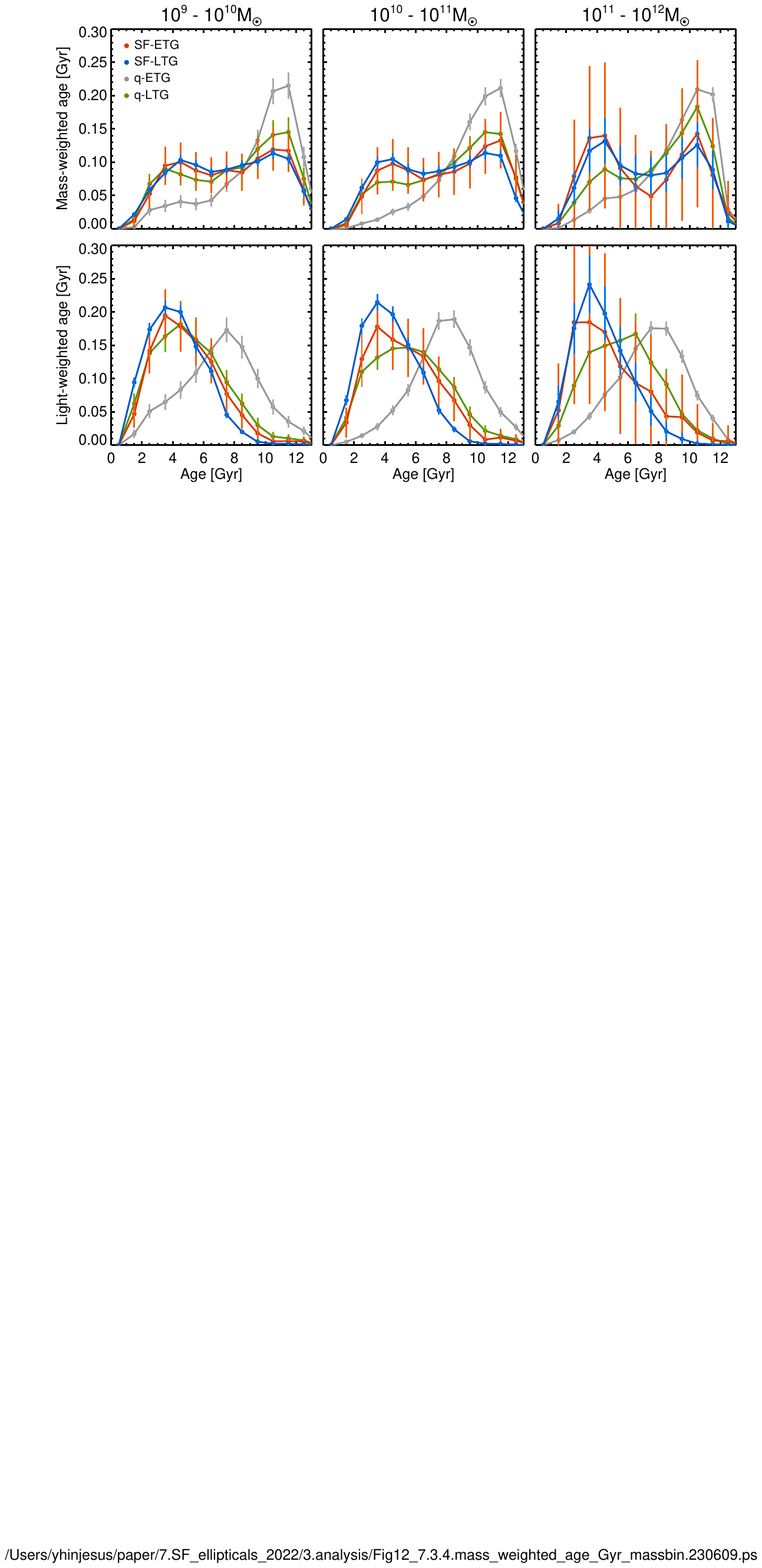}
\caption{The fractions of stellar population with mass-weighted (top) and light-weighted (bottom) age, divided by three mass bins. These fractions are measured in the same way as in Figure \ref{Fig2}. The errors are measured by the confidence interval of the fraction for a beta distribution, which depends on the sample size.}\label{Fig12}
\end{figure*}

\end{document}